\documentclass[12pt]{iopart}

\expandafter\let\csname equation*\endcsname\relax
\expandafter\let\csname endequation*\endcsname\relax
\usepackage{amsmath}
\usepackage[final,nopatch=footnote]{microtype}
\usepackage[utf8]{inputenc}                             
\usepackage[T1]{fontenc}                                
\usepackage{hyperref}                                   
\usepackage{url}                                        
\usepackage{booktabs}                                   
\usepackage[numbers,sort&compress]{natbib}
\usepackage{amsfonts,amssymb}                           
\usepackage{bm,graphicx}
\usepackage{microtype}                                  
\usepackage{xcolor}
\usepackage{setspace} 
\usepackage{array}
\usepackage{etoolbox}

\def\bea{\begin{eqnarray}}
\def\eea{\end{eqnarray}}
\def\ba{\begin{array}}
\def\ea{\end{array}}

\def\la{\langle}
\def\ra{\rangle}

\makeatletter
\newrobustcmd{\fixappendix}{%
  \patchcmd{\l@section}{1.5em}{7em}{}{}%
  \patchcmd{\l@subsection}{2.3em}{7em}{}{}%
}
\makeatother

\makeatletter
\newcommand{\reqnomode}{\tagsleft@true}
\newcommand{\leqnomode}{\tagsleft@false}
\makeatother

\makeatletter
\def\@mkboth#1#2{}
\newlength\appendixwidth
\preto\appendix{\addtocontents{toc}{\protect\patchl@section}}
\newcommand{\patchl@section}{%
  \settowidth{\appendixwidth}{\textbf{Appendix }}%
  \addtolength{\appendixwidth}{1.5em}%
  \patchcmd{\l@section}{1.5em}{\appendixwidth}{}{\ddt}%
}
\makeatother

\begin{document}

\title{Tracer dynamics in an interacting active bath: fluctuations and energy partition}
\author{Ritwick Sarkar}
\address{S. N. Bose National Centre for Basic Sciences, Kolkata 700106, India}
\author{Ion Santra\footnote{ion.santra@theorie.physik.uni-goettingen.de}}
\address{Institut für Theoretische Physik, Georg-August-Universität Göttingen, Goettingen 37077,  Germany}
\begin{abstract}
We investigate the dynamics of a massive tracer particle coupled to an interacting active bath, modeled as a harmonic chain of overdamped active particles analytically, with an aim to understand the impact of bath interactions and activity on the nonequilibrium fluctuations of the tracer. From the microscopic equations, we derive the tracer particle's effective Langevin equation, obtaining the dissipative and stochastic forces from the bath. We analyze the friction kernel, revealing power-law tails in the weak coupling limit and exponential decay in the strong coupling regime. Due to the interplay between bath interactions, probe-bath coupling, and activity, the mean squared displacement, velocity, and stationary velocity correlations exhibit different dynamical regimes, which we characterize analytically. Under harmonic confinement, we find that energy equipartition holds at low activity but breaks down at higher activity, with the kinetic energy exhibiting a non-monotonic dependence on the activity of the bath. 
\end{abstract}

\maketitle
\tableofcontents
\section{Introduction}
Coarse-grained stochastic descriptions have been remarkably successful in capturing the dynamics of suspended particles in fluids. A paradigmatic example is Brownian motion, where a colloidal particle undergoes random motion due to collisions with surrounding fluid molecules—a phenomenon elegantly described by the Langevin equation~\cite{langevin1908}. In this framework, the influence of the medium is modeled through two essential components: a deterministic dissipative force and a random thermal force. At equilibrium, these two terms are constrained by the fluctuation-dissipation theorem (FDT)~\cite{Kubo}, ensuring consistency with the equilibrium statistical mechanics. Initially introduced as a phenomenological model, Langevin equations have since been derived from microscopic Hamiltonian dynamics using exact calculations, projection operator techniques~\cite{zwanzig1973nonlinear,zwanzigbook}. These derivations yield explicit forms for dissipation and noise while preserving the fundamental connection dictated by FDT.

 Nonequilibrium media are ubiquitous in complex systems, encompassing examples such as active particle suspensions~\cite{active_suspension1,active_suspension2}, glassy materials~\cite{noneq_glass1,noneq_glass2}, sheared fluids~\cite{sheared_noneq1,sheared_noneq2}, and intracellular environments~\cite{noneq_intracellular1}. In these systems, the equilibrium principles outlined in the previous paragraph no longer hold: there is no universal framework akin to FDT, and the behavior of a tracer depends sensitively on the underlying microscopic dynamics. Tracers in nonequilibrium environments exhibit a strikingly rich set of behaviors, often in stark contrast to their equilibrium counterparts. These include modification of fluctuation-dissipation relation~\cite{seyforth2022nonequilibrium,boriskovsky2024fluctuation,wu,maes_activebath,santra2023dynamical,khali2023does,Solon_2022,tracer_diff1,maggi2017memory} and equipartition theorem \cite{maggi_activebath}, emergence of nonequilibrium forces and relaxations~\cite{shea2022passive,force_renormalization,santra2024forces,knotz2024mean}, and memory~\cite{caprini2024emergent},
anomalous transport and non-Gaussian fluctuations \cite{micromotor,greneck_activebath,transport_bacterial_turbulence,dhar2024active,Knezevic_2020,Bello2024,Put_2019,hargus2025passive}, unusual thermodynamic properties \cite{stirling_engine,D3SM01177A,tanogami2022violation,venturelli2024stochastic}, effective interaction among tracer particles ~\cite{maggi_2,tracer_diff2}, generation of active fluctuations~\cite{pei2025transfer,al2025cold1}, anomalous energy transport~\cite{santra2022activity,steady_harmonics_chain,prsa_rubin}.

To understand these phenomena theoretically, an open system framework is being employed in the recent years, where probe dynamics coupled to a nonequilibrium medium is considered, the latter consisting of degrees of freedom that break detailed balance. In this context, stochastic models of active particle dynamics, which have been quite  successful in understanding the rich physics of active matter, have emerged as one of the leading candidates. \textcolor{black}{The approach is in the same spirit as the open-system framework used to study quantum Brownian motion~\cite{ambegaokar1986quantum,ford1988quantum,das2020quantum} and mimics the famous experiment by Wu et al.~\cite{wu}, which reported anomalous diffusion of colloidal beads in an E. coli suspension.} Such microscopic description provides great insights into what
roles the parameters of the fluctuating nonequilibrium environment and coupling have on the dynamics of the probe. One of the simplest ways of getting analytical insights is by studying non-interacting baths similar to the famous \textcolor{black}{Caldeira-Leggett} model~\cite{Caldeira_Leggett_1} revealing exact forms of modified FDTs and response relations~\cite{santra2023dynamical}, emergence of negative friction~\cite{maes_activebath,pei2025transfer}. However, analytical studies on dynamics of a massive tracer in interacting baths have been limited~\cite{pei2025transfer}.

\textcolor{black}{In this paper, we aim to get a comprehensive analytical understanding of tracer dynamics in an interacting active bath, with particular focus on how the competition between the bath interactions and the tracer-bath coupling affects the dynamics. To
this end, we model} a massive probe in an interacting active bath modeled by a harmonic chain of overdamped active particles. We derive the effective Langevin equation of the probe, deriving the exact forms of the dissipative and stochastic forces coming from the bath and discuss the strong and weak coupling limits. In particular, we show that the friction kernel in the weak coupling limit is characterized by power-law tails, while the strong coupling limit can be effectively described by faster exponential tails. We then explore the mean squared displacement and velocity of the probe, without any external confinement, which show rich dynamical behavior that emerges as an interplay of the bath interaction, coupling and activity. We compute the stationary velocity correlations of the probe which are characterized by a short-time exponential decay, and a long time power-law decay. We then explore the equipartition energy, by computing the kinetic and potential energies of the probe in an external harmonic confinement--- while equipartition of energy holds for small activity with an effective temperature, it is violated for larger values of activity. Moreover, both kinetic and potential energies depend on the strength of the external confinement, with the kinetic energy displaying a non-monotonic dependence on the activity of the medium. 

The paper is organized as follows: In Sec.~\ref{sec_structure}, we introduce the model and derive exact expressions for the dissipation kernel and autocorrelation of the noise experienced by the tracer; we also discuss the modified FDT. In Sec.~\ref{dyn_fluct}, we explore the mean squared velocity and displacement of the tracer, and report the temporal behavior in the different dynamical regimes. In Sec.~\ref{autocorrelation}, we study the relation between the response to an external perturbation and the velocity autocorrelation function of the tracer. Thereafter, in Sec.~\ref{energetics} we put the tracer in an additional harmonic trap and investigate the partition of the kinetic and potential energies.

\section{Model}\label{sec_structure}

We investigate the dynamics of a particle of mass $m$ (probe) coupled to the end particle of an extended, interacting active reservoir, \textcolor{black}{named `active Rubin bath' (ARB)}~\cite{prsa_rubin} via harmonic spring of stiffness $\lambda$. The active reservoir is modeled by $M$ overdamped active oscillators coupled to its nearest neighbour with a harmonic spring of stiffness constant $k$. The position of the probe, denoted by $x$, evolves by,
\begin{align}
    m\ddot{x}(t)=-\lambda(x-y_M)-\partial_x U(x)\label{eq:x}
\end{align}
where $U(x)$ is an external potential experienced by the tracer; $y_M$ denotes the coordinate of the end of the active reservoir that evolves by the Langevin equation,
\begin{align}
    \nu\dot{y}_M=k[y_{M-1}(t)-2 y_M(t)]+\lambda x(t)-(\lambda-k)y_M(t)+f_M(t).\label{eom:bath1}
\end{align}
The other end of the active reservoir is connected to a fixed wall, 
\begin{align}
\nu\dot{y}_l=
        k[y_{l-1}(t)+y_{l+1}(t)-2 y_l(t)]+f_l(t)\quad \forall~~l \in [1,M-1]\label{eom:bathl}
\end{align}
with $y_{0}=0$. \textcolor{black}{Here $\nu$ is the friction coefficient and $f_l(t)$ is the active force acting on the $l$-th particle.} The assumption of harmonic coupling in this model should be viewed as a mathematical simplification rather than a strictly physical constraint. In fact, if we consider a two body interaction potential $V(|z_1-z_2|)$ with a short-range, the leading order term in the Taylor expansion of the potential around its equilibrium separation $z_{\text{eq}}$ gives an effective measure of the spring constants $k=V''(z_{\text{eq}})$ [and similarly, $\lambda$]. \textcolor{black}{It is worth mentioning that the dynamics of ARB follow equations of motion similar to an overdamped Rouse polymer chain, albeit with an active noise. In fact, the effective generalized Langevin equations(GLE) or fractional Langevin equations for tagged monomers in overdamped Rouse polymer chains have been studied before in the context of Markovian embedding~\cite{goychuk2012viscoelastic}, anomalous diffusion \cite{gle_polymer,gle_anomalous2,fLE_1,tagged_bead,anomalous_diffusion,gle_rouse_chain}, response to nonequilibrium drivings~\cite{Maes_rouse,gle_probe_bath}.  This has also been studied for tagged monomers in active chains which show further interesting behavior \cite{gle_probe_bath, Singh_2021, Prakash_2024, D4SM00350K}.}

\begin{figure}[t]
    \centering
\includegraphics[width=0.5\linewidth]{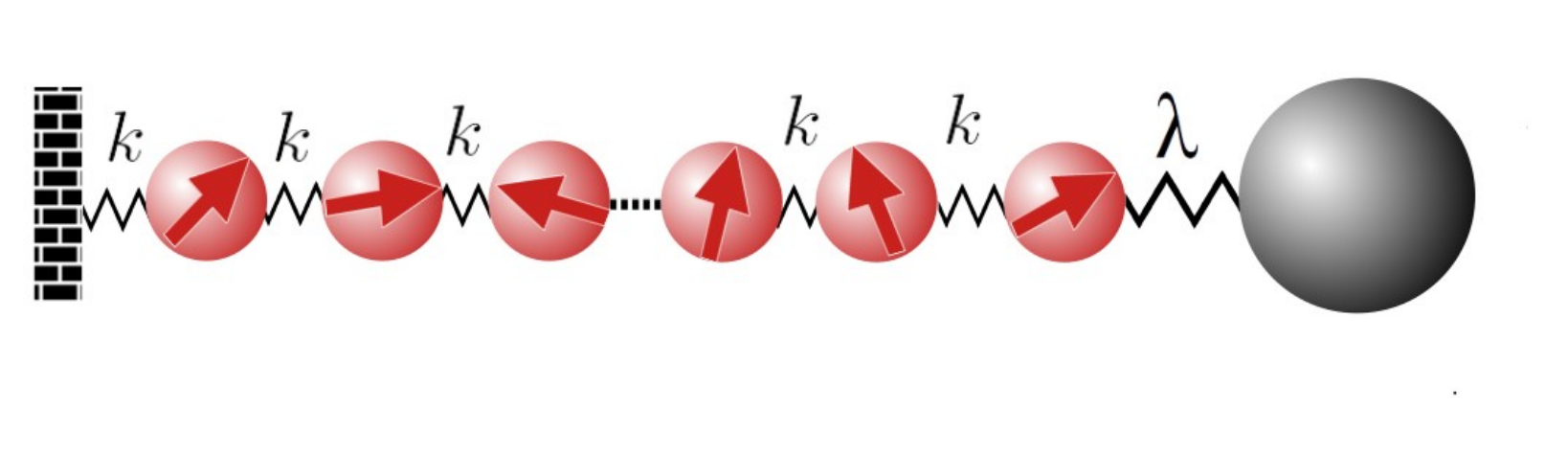}
    \caption{Schematic description of \textcolor{black}{a massive probe attached to an active Rubin bath. The probe is shown in gray, while the active bath particles are represented as red circles with arrows.}}
    \label{fig:model}
\end{figure}
Each bath particle is driven by an independent active force $f_l(t)$, which is stochastic in origin with zero mean and \textcolor{black}{a} correlation,
\begin{align}
    \la f_l(t)f_{l'}(0)\ra=h(t,\tau)\delta_{ll'}.
    \label{fcorr}
\end{align}
The specific form of $h(t,\tau)$ depends on the dynamics of the active force driving the bath particle. In this paper, we model the bath particles as one-dimensional run-and-tumble particles: $f(t)$ is a dichotomous stochastic variable that switches between \textcolor{black}{$\pm v_0$} at a rate $(2\tau)^{-1}$ \cite{Malakar_2018,romanczuk2012active,santra2020run}. This leads to,
\begin{align}
    h(t)=v_0^2e^{-t/\tau}.\label{fcorr2}
\end{align}
Note that, this exponentially decaying form of the autocorrelation function is a generic feature of the well-known active particle models~\cite{bechinger2016active,santra2022universal} and can  
 microscopic stochastic dynamics as well: e.g., active Ornstein-Uhlenbeck particles \cite{aoup_1,aoup_2}, and active Brownian and direction reversing active Brownian particles \cite{ABP1,ABP2,santra2021active}. The correlation time $\tau$ determines how active the constituents are, and is considered as the measure for activity.

The bath equations of motion can be recast in a matrix notation, 
\begin{align}
\dot{Y}(t)&=\Phi Y(t)+W X(t)+F(t),\label{yeqm}
\end{align}
where $Y^T(t)=\{y_1(t),y_2(t)\cdots y_M(t)\}$, $X^T(t)=\{0 \cdots x(t)\}$, $F^T=\{f_1(t)/\nu ,\cdots f_M(t)/\nu\}$, and $W_{ij}=(\lambda/\nu) \delta_{iM} \delta_{jM}$. The force matrix $\Phi$ is a tridiagonal matrix with elements,
\begin{align}
    \Phi_{ij}=\begin{cases}
         -\frac{k}{\nu}(2\delta_{ij}-\delta_{i~j-1}-\delta_{i~j+1})~~\text{for }i,j\neq M\\-\frac{(\lambda+k)}{\nu}\delta_{iM}\delta_{jM}       
    \end{cases}
\end{align}

To integrate out the bath degrees of freedom, we first look for a solution of Eq.~\eqref{yeqm} for a given tracer position $x(t)$,
\textcolor{black}{
\bea
{Y}(t) = \int_{-\infty}^t ds   \Big[ \mathcal{U}^{-1} e^{D\frac{(t-s)}{\nu}} \mathcal{U} W X(s)+\mathcal{U}^{-1} e^{D\frac{(t-s)}{\nu}}\mathcal{U} F(s)\Big].\label{eom_matrix_active}
\eea
}
where $D$ denotes the diagonalized matrix of $\Phi$, defined by the similarity transformation $D=\mathcal{U} \Phi \mathcal{U}^{-1}$. The solution for the $M$-th bath particle can be obtained using the explicit forms of $W$ and $X(t)$,
\begin{align}
     y_M(t)&=\int_{-\infty}^t~ds\,\left[\frac{\lambda}{\nu}\Lambda_{MM}(t-s)x(s)+\frac{1}{\nu}\Lambda_{Mj} \textcolor{black}{(t-s)} f_j(s)\right],\label{temp_lambda1}
\end{align}
\textcolor{black}{here repeated indices indicate a summation over them and $\Lambda=\mathcal{U}^{-1} e^{D\frac{(t-s)}{\nu}} \mathcal{U}$ can be interpreted as the Green's function matrix}. Inserting Eq.~\eqref{temp_lambda1} in Eq.~\eqref{eq:x}, we get for the probe position,
\begin{align}
 m\ddot{x}(t)&=-U'(x)- \int_{-\infty}^t ds\Gamma(t-s)\dot{x}(s) +\zeta(t).
 \label{eom_tracer}
 \end{align}
\textcolor{black}{The above equation is the generalized Langevin equation(GLE) for the massive probe in a harmonic trap and also attached to an active Rubin bath.} The effective noise $\zeta(t)$ and dissipation kernel $\gamma(t)$ experienced by the probe are defined by,
\begin{align}
 \zeta(t) &=\frac{\lambda}{\nu}\int_{-\infty}^t ds\, \Lambda_{Mj}(t-s)f_j(s),\label{eq:zeta1}\\
  \textcolor{black}{\frac{\lambda^2}{\nu}} \Lambda_{MM}(t) &=-\frac{d}{dt}\Gamma(t) ~~\text{with } \Gamma(t)=\gamma(t)\Theta(t),\label{gammat}
\end{align}
characterize the forces coming from the active reservoir. Here $\Theta(z)$ is the Heaviside theta function that preserves the causal nature of the dissipation kernel \cite{DLMF}.

To understand these forces, we need to compute the elements of the Green's function matrix $\Lambda$. This can be done by diagonalizing the $\Phi$ which involves the task of computing the eigenfunctions of $\Phi$ which is quite cumbersome. Instead we use the results for a simpler system $\lambda=k$ [henceforth we will refer to it as \emph{pure system}] which was computed recently~\cite{prsa_rubin} to obtain the necessary matrix elements of $\Lambda$. To this end, we first note that the Green's function in frequency domain is,
\begin{align}
    \tilde{\Lambda}(\omega)=[-i \omega \mathbb{I}-\Phi]^{-1}.\label{gf1}
\end{align}
where tilde denotes the Fourier transform defined by $\tilde f(\omega)=\int_{-\infty}^{\infty}dt e^{i\omega t} f(t)$, and $\mathbb{I}$ denotes the identity matrix. This can be expressed in terms of the Green's function of the \emph{pure system} $\tilde{\Lambda}^{0}(\omega)$ as,
\begin{align}
   \tilde{\Lambda}(\omega)=\left([\Lambda^{0}]^{-1}+\Phi'\right)^{-1} ,\quad\text{with }  \tilde{\Lambda}^{0}(\omega)=[-i \omega \mathbb{I}-\Phi^{0}]^{-1},\label{lambda_1}
\end{align}
where $\Phi^0$ denotes the force matrix with $\lambda=k$, and $\Phi'_{ij}=\frac{(\lambda-k)}{\nu}\delta_{iM}\delta_{jM}$. Using the explicit form of $\Phi'_{ij}$, the above equation can be simplified further to obtain,
\begin{align}
     \nu \tilde\Lambda_{ij}(\omega)+(\lambda-k)\tilde\Lambda^{0}_{iM}(\omega)\tilde\Lambda_{Mj}(\omega) = \nu\tilde\Lambda^{0}_{ij}(\omega). \label{Component_greens}
\end{align}
In the following we use this relation to obtain closed analytical forms for the damping kernel and effective noise correlation.

\subsection{Dissipation kernel}
Let us first discuss the dissipation kernel $\tilde\gamma(\omega)$, defined in Eq.~\eqref{gammat}. This requires computation of the matrix element $\tilde\Lambda_{MM}(\omega)$ only; and putting $i=j=M$ in Eq.~\eqref{Component_greens}, we arrive at a simple relation for $\tilde\Lambda_{MM}$,
\begin{align}
    \tilde\Lambda_{MM}(\omega)=\tilde\Lambda_{MM}^0(\omega)\left [\mathbb{I}+\frac{\lambda-k}{\nu}\tilde\Lambda_{MM}^0(\omega)\right]^{-1}
\end{align}
The real and imaginary parts of dissipation kernel $\tilde\gamma(\omega)$, denoted by $\tilde\gamma'(\omega)$ and $\tilde\gamma''(\omega)$, respectively, can then be obtained using Fourier transform of Eq.~\eqref{gammat} as,
\begin{align}
\omega\,\tilde\gamma'(\omega) = \frac{\lambda^2}{\nu}\tilde \Lambda_{MM}''(\omega),\quad
\omega\,\tilde\gamma''(\omega)=\gamma(0) -\frac{\lambda^2}{\nu}\tilde \Lambda_{MM}'(\omega),\label{gamma-omega-identity}
\end{align}
where $\gamma(t=0)=\frac{2}{\pi}\int_0^\infty d\omega \tilde\gamma'(\omega)$.
In the thermodynamic limit of a large number of bath oscillators, we use $\tilde\Lambda_{MM}^{0}(\omega)$ from Ref.~\cite{prsa_rubin}, to obtain the real and imaginary parts of the dissipation kernel (see also \textcolor{black}{\ref{app_pure}}),
\begin{align}
\tilde\gamma'(\omega)&=\frac{2 k^2 \lambda ^2 \nu  p q^2}{4 k^2 (k+\lambda  p)^2+\nu ^2 p^2 q^2 \omega ^2 (k-\lambda )^2}\\
\tilde\gamma''(\omega)&=\frac{\lambda}{\omega}\left( 1 - \frac{\lambda  p \left(4 k^2 (k+\lambda  p)+\nu ^2 p q^2 \omega ^2 (\lambda -k)\right)}{4 k^2 (k+\lambda  p)^2+\nu ^2 p^2 q^2 \omega ^2 (k-\lambda )^2}\right)
\end{align}
with $q=\sqrt{\frac{1+\sqrt{1+16k^2/(\nu^2\omega^2)}}{2}}$, and $q=p-1$. Note that, the $\tilde\gamma'(\omega)$ and $\tilde\gamma''(\omega)$ are even and odd function of $\omega$ respectively.

At this stage it is important to identify the three time-scales involved: (i) the active time-scale $\tau$, which characterizes the memory in the stochastic forces of the bath oscillators, (ii) $\tau_b=\nu/k$ which is the characteristic time scale of the bath oscillators, and denotes the time required for the bath oscillators to reach mechanical equilibrium, and (iii) the probe-bath coupling time-scale $\tau_c=\nu/\lambda$. It is important to note that the dissipation is independent of the nature of the active force driving the bath particles, which is a direct consequence of the harmonic nature of the couplings in the system. 

For $\tau_b<\tau_c$, both the real and imaginary parts of $\tilde\gamma(\omega)$ have the same decay for small $\omega\ll (\tau_b^{-1},\tau_c^{-1})$,
\begin{align}
   \tilde\gamma'(\omega\to 0^+)=\tilde\gamma''(\omega\to 0^+) \approx \sqrt{\frac{\nu k}{2}}\omega^{-1/2}
\end{align}
which is independent of the probe bath coupling $\lambda$. Thus at longer time intervals, the memory in the system depends solely on the structure of the bath. It is interesting to point out that the imaginary part of the memory kernel has a discontinuity at $\omega=0$: it diverges to $\pm \infty$ as $\pm\omega^{-1/2}$ as $\omega$ approaches zero from the positive and negative sides, respectively. This is different than the single bath particle case, or the case of underdamped bath particles (as in the original work of Rubin\textcolor{black}{~\cite{rubin_bath}}), where the imaginary part of the memory kernel approaches $0$ from the positive and negative sides. 

 For large frequencies, $\tilde\gamma'(\omega)$ decreases as $\omega^{-2}$, while $\tilde\gamma''(\omega)$ decreases as $\omega^{-1}$,
 \begin{align}
     \tilde\gamma'(\omega\to\infty)\approx \frac{\lambda^2}{\nu \omega^2}, \quad\text{and ~~}    \tilde\gamma''(\omega\to\infty)\approx \frac{\lambda}{\omega}
 \end{align}
which is independent of the structure of the bath interaction $k$. This is expected since at very short-time intervals the probe effectively sees only the end particle of the bath to which it is coupled to. 

 The small and large frequency dependences of the dissipation kernel is the same as previously seen in Ref~\cite{prsa_rubin}, albeit with modified prefactors. An interesting intermediate regime $\tau_c^{-1}\ll \omega\ll \tau_b^{-1}$ emerges due to the different relaxation times of the chain and the coupling. In this regime $\tilde\gamma'(\omega)$ becomes independent of $\omega$, while \textcolor{black}{$\tilde\gamma''(\omega)$} increases as $\omega$.
\begin{figure}
    \centering
    \includegraphics[width=0.85\linewidth]{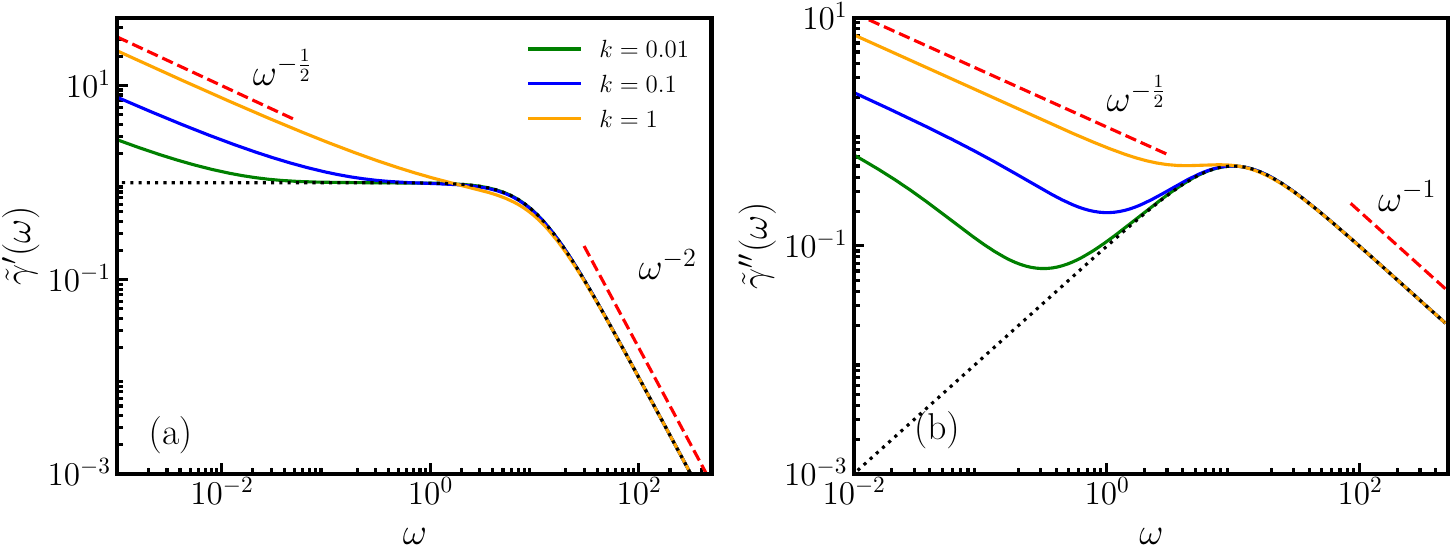}
    \caption{Panels (a) and (b) show the real and imaginary parts of $\tilde\gamma(\omega)$, respectively, for $\lambda=10$, $\nu=1$, and different values of $k$. The dashed lines in both panels represent  the real and imaginary parts of $\tilde\gamma(\omega)$ an exponential dissipation kernel.}
    \label{f:gamma-kernel}
\end{figure}
The small frequency $\omega^{-1/2}$ behavior implies a large-time $t^{-1/2}$ decay of the memory kernel. This implies that the integral $\int_0^\infty dt'  \gamma(t')$ diverges, meaning that a time-scale separation among the tracer and the bath is not a good approximation for this kind of model. However, for $\tau_b\ll \tau_c$, the initial regime becomes smaller and smaller, and $\tilde\gamma(\omega)$ almost resembles a Lorentzian, indicating an exponential form for the dissipation kernel, $\gamma(t)\approx e^{-t/\tau_c}$ [shown in both panels of Fig.~\fref{f:gamma-kernel} with dashed lines].
Physically this is the scenario when the bath oscillators are weakly coupled to each other while the probe bath coupling is very strong. Additionally, in this regime when $\tau_c$ is very small, then the memory decays very fast, and an effective Markovian description emerges. \textcolor{black}{The slow power-law decay of the memory kernel at long times can be physically understood by noting that any local strain induced by the tracer~\cite{gle_polymer,gle_anomalous2} takes an infinite amount of time to fully propagate through the infinite bath. This leads to a long-lived back-action on the tracer. The power-law exponent of the decay arises from the long-range harmonic interactions in the bath and matches that of the memory kernel experienced by a tagged monomer in an infinite Rouse chain. As we will see later, this long memory leads to subdiffusive dynamics of the tracer at late times.}

\subsection{Effective noise correlations}
\begin{figure}
    \centering
    \includegraphics[width=0.55\linewidth]{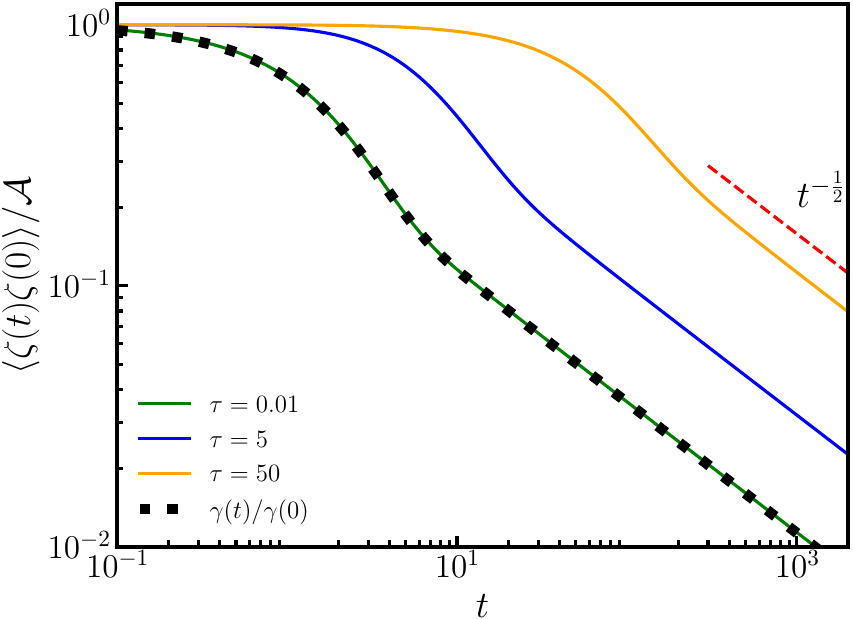}
    \caption{Effective noise correlations of the tracer \textcolor{black}{attached to an ARB} for $k=1,\,\lambda=5,\,\nu=10$, and $\mathcal{A}=\la \zeta^2(0)\ra$.}
    \label{noise:corr:fdt}
\end{figure}

The autocorrelation of the effective noise $\zeta(t)$, defined in Eq.~\eqref{eq:zeta1} can be computed using the Eq.~\eqref{Component_greens} as,
\begin{align}
    \la \zeta(t)\zeta(t')\ra=\frac{\lambda^2}{\nu^2}\int_{-\infty}^t ds_1\,\int_{-\infty}^{t'} ds_2 \Lambda_{Mi}(t-s_1)\Lambda_{Mj}(t-s_2)\la f_i(s_1)f_j(s_2)\ra,\label{noise_time}
\end{align}
here repeated indices indicate a summation over them. The independence of the noise Eq.~\eqref{fcorr} reduces the double sum in the above equation to a single one, which can again be converted to an integral in the thermodynamic limit $M\to\infty$. Similar to the dissipation kernel, it turns out, that it is more convenient to calculate the noise correlation in the frequency domain, and from Eq.~\eqref{noise_time}, we have,
\begin{align}
    \left\langle \tilde\zeta(\omega)\tilde\zeta(\omega')  \right\rangle=\frac{\lambda^2}{\nu^2}  \tilde\Lambda_{Mi}(\omega)\tilde\Lambda_{Mj}(\omega') \langle\tilde{f}_i(\omega) \tilde{f}_{j}(\omega')\rangle.\label{noise_freq}
\end{align}
To evaluate the rhs of the above equation, we need the matrix elements $\tilde{\Lambda}_{Mj}(\omega)$ as well as the noise correlation $\langle\tilde{f}_i(\omega) \tilde{f}_{j}(\omega')\rangle$. To this end, putting $i=M$ in Eq.~\eqref{Component_greens}, we get,
\begin{align}
    \tilde\Lambda_{Mj}(\omega)=\tilde\Lambda_{Mj}^0(\omega)\left [\mathbb{I}+\frac{\lambda-k}{\nu}\tilde\Lambda_{MM}^0(\omega)\right]^{-1}.\label{matrix_l1}
\end{align} 
Further, in frequency space the noise correlations Eq.~\eqref{fcorr} become,
\begin{align}
    \langle\tilde{f}_i(\omega) \tilde{f}_{j}(\omega')\textcolor{black}{\ra}=\delta_{ij}\tilde h(\omega,\tau)=\delta_{ij}\frac{2 v_0^2 \tau}{1+\omega^2 \tau^2}.\label{noise_1}
\end{align}
Using Eqs.~\eqref{matrix_l1} and \eqref{noise_1} in Eq.~\eqref{noise_freq} and simplifying using the relation given in Eq.~\eqref{gamma-omega-identity} we finally arrive at, 
\begin{align}
   \left\langle \tilde\zeta(\omega)\tilde\zeta(\omega')  \right\rangle =\frac{2 \pi}{\nu} \tilde{\gamma}'(\omega)\tilde{h}(\omega,\tau)\delta(\omega+\omega')\label{noise:autocorr}
\end{align}
This is the modified Fluctuation dissipation theorem (FDT) for the probe. The ratio of the noise autocorrelations and real part of the dissipation kernel is given by a frequency dependent function, unlike in equilibrium where it is proportional to the bath temperature. Only in the limit of small activity $\tau\to 0$, $\tilde h(\omega,\tau)\approx 2v_0^2\tau$, an effective thermal behavior is recovered with an effective temperature $v_0^2\tau/\nu$. This is showed in Fig.~\ref{noise:corr:fdt} where the normalized noise correlation and friction kernel \textcolor{black}{are} plotted in time--- an overlap of the two for small $\tau$ illustrates the effective thermal behavior. 

\section{Dynamical fluctuations: mean squared displacement and velocity}\label{dyn_fluct}
\subsection{Mean squared displacement}
The dissipation kernel and noise correlations, though, capture the influence of the active medium on the probe particle, they are not directly measurable in experiments. Instead, the mean squared displacement (MSD) and mean squared velocity (MSV) serve as two widely used observables that bridge theoretical predictions and experimental observations. These quantities can be experimentally accessed through techniques such as single-particle tracking and dynamic light scattering. Here, we focus on an unconfined probe, i.e., $U(x)=0$ [see Eq.~\eqref{eq:x}], and compute its MSD and MSV.

The MSD in the stationary state can is given by,
\begin{align}
     \Delta^2_x(t) =\la (x(t)-x(0))^2\ra=2\la x^2(t)-x(t)x(0)\ra 
    =\frac{2}{\nu \pi}\int_0^\infty d\omega \frac{\tilde h(\omega,\tau)\tilde \gamma'(\omega)(1-\cos{\omega t})}{[-m \omega^2+\omega \tilde\gamma''(\omega)]^2+\omega^2\tilde\gamma'^2(\omega)}. \label{msd_numerical}
\end{align}
To get the final integral form above, we have used the solution of Eq.~\eqref{eom_tracer} in frequency space,
\begin{align}
    \tilde x(\omega)=\frac{\tilde\zeta(\omega)}{-m\omega^2-i\omega\tilde\gamma(\omega)}
    \label{eq:xomega}
\end{align}
and the form of noise correlations given by Eq.~\eqref{noise:autocorr}. Though the above integral is very hard to compute exactly, it can be computed numerically, and in the following, we discuss them for the different parameter regimes.

\subsubsection{Slower bath relaxation \texorpdfstring{$\tau_c<\tau_b$}{}}\label{subsection1msd}
\begin{figure}[ht]
    \centering
    \includegraphics[width=\linewidth]{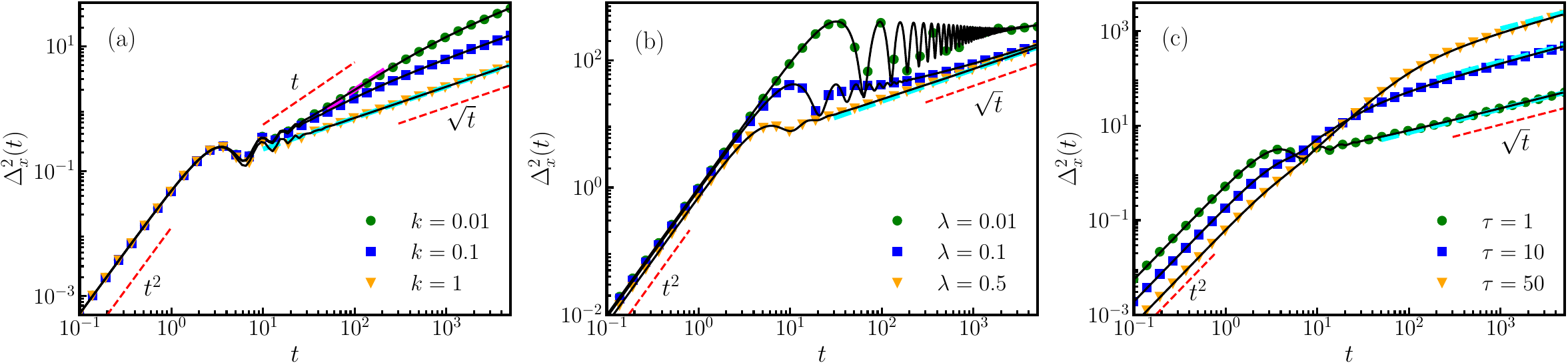}
    \caption{Mean square displacement (MSD) of the probe particle attached to an \textcolor{black}{ARB} as a function of time showing the MSD in the (a) Strong coupling regime: for different values of $k$ and other parameters fixed at $\nu=10$, $\lambda=1$ , $\tau=1$; and (b) Weak coupling regime: for different values of $\lambda$ and other parameters fixed at $\nu=1$, $k=1$, $\tau=1$. Panel (c) shows the  MSD for different values of the activity $\tau$ and other parameters fixed at $\nu=1$, $k=10$, $\lambda=1$. Black solid lines and cyan dashed lines correspond to Eqs.~\eqref{msd_numerical} and \eqref{sqrrd_t} respectively. The magenta dashed line in panel (a) correspond to Eq.~\eqref{lin_t}. \textcolor{black}{Numerical simulations are averaged over $10^4$ trajectories with $\Delta t=5\times 10^{-3}$.}}
    \label{fig:msd}
\end{figure}

Let us first discuss the temporal behavior of the MSD for $\tau_c<\tau_b$, i.e., $\lambda>k$. In this parameter regime, the probe behaves as if it is coupled to a single bath particle till $t\ll \tau_b$. At very short-times, $t\ll \tau_c$, the probe shows a ballistic behavior which continues in the regime $\tau_c\ll t\ll\tau$. The leading order asymptotic behavior can be obtained as,
\begin{align}
     \Delta^2_x(t)\sim A t^2\quad\text{with }A=\frac{1}{\nu \pi}\int_0^\infty d\omega \frac{\omega^2\tilde h(\omega,\tau)\tilde \gamma'(\omega)}{(-m \omega^2+\omega \tilde\gamma''(\omega))^2+\omega^2\tilde\gamma'^2(\omega)}.
    \label{msd:ballistic}
\end{align}
This ballistic behavior crosses over to a diffusive behavior $\sim t$ for $\tau\ll t\ll \tau_b$. \textcolor{black}{This can be understood as follows: the probe and the first bath particle attains a mechanical equilibrium at time $\sim \tau_c$, whereafter, the probe follows the motion of the first bath particle. The bath particle, which itself follows run-and-tumble dynamics, has a ballistic \textcolor{black}{MSD} for $t\ll \tau$ and diffusive for $t\gg\tau$ resulting in the above mentioned behavior of MSD.}  The leading order diffusive behavior is independent of the $\lambda$ and $k$, and asymptotically given by,
\begin{align}
     \Delta^2_x(t)\simeq \frac{2v_0^2\tau}{\nu^2}t. \label{lin_t}
\end{align}
\textcolor{black}{
These predictions are compared to numerical simulations of Eqs.~\eqref{eq:x}-\eqref{eom:bathl} [see \ref{app_numerucal} for details on numerical simulations] in  Fig.~\ref{fig:msd}(a), and show excellent agreement.}
Thereafter, beyond $t\simeq\tau_b$, the probe feels the effect of all the bath particles, and the MSD shows a subdiffusive behavior. The leading order asymptotic behavior can be extracted from Eq.~\eqref{msd_numerical} as,
 \begin{align}
    \Delta^2_x(t)\approx \frac{ 4 v_0^2\tau}{\sqrt{\pi k } \nu ^{3/2}}\sqrt{t}.\label{sqrrd_t}
\end{align}
\textcolor{black}{This subdiffusive behavior at long times is because any displacement in the tracer requires a cumulative response from the ARB, which relaxes as a slow power-law at long times.}

\subsubsection{Faster bath relaxation \texorpdfstring{$\tau_c>\tau_b$}{}}
On the other hand, for $k>\lambda$, i.e., $\tau_b<\tau_c$, i.e., the bath oscillators relaxes to a mechanical equilibrium faster than the coupling with the probe. Let us consider the case when $\tau$ is the smallest time-scale: the probe shows the initial ballistic regime which crosses over to a $\sim\sqrt{t}$ behavior at $t\sim \tau_c$ [see Fig.~\ref{fig:msd}(b)]. The behavior remains the same when $\tau$ is intermediate between $\tau_b$ and $\tau_c$. When however, $\tau$ is the largest time-scale among the three then, the crossover between the ballistic to sub-diffusive $\sim\sqrt{t}$ behavior happens at $t=\tau$, same as Eq.~\eqref{sqrrd_t}. \textcolor{black}{The physical origin of the growth exponents are the same as discussed for the previous subsection~\ref{subsection1msd}.}

\subsubsection{Oscillations in MSD}
For all the cases discussed above, we find that at the onset of the intermediate- or late-time regimes, we see damped oscillatory behavior about the power-law behaviors. The oscillations, of frequency $\sqrt{\lambda/m}$, occur due to the restoring force felt by the probe from the first bath oscillator. Naturally, the amplitude of the oscillations depends on the relative strength of the inter-bath and probe-bath interactions: it is more pronounced in the weak coupling limit, and decreases with the increase in the ratio $\lambda/k$ for a fixed activity [see Fig.~\ref{fig:msd}(b)]. Interestingly, the oscillations also decrease with increasing bath activity $\tau$ [see Fig.~\ref{fig:msd}(c)].  Table~\ref{tab:msd_msv} summarizes the temporal behavior of the MSD in the different different dynamical regime.

\subsection{Mean squared velocity}
The probe velocity $v(t)=\dot{x}(t)$ in Fourier space is easily obtained as $\tilde v(\omega)=-i\omega \tilde x(\omega)$. Thus, the MSV is given by,
\begin{align}
     \Delta^2_v(t) =2\la v^2(t)-v(t)v(0)\ra
    &=\frac{2}{\nu \pi}\int_0^\infty d\omega \frac{\omega^2\,\tilde h(\omega,\tau)\tilde \gamma'(\omega)(1-\cos{\omega t})}{(-m \omega^2+\omega \tilde\gamma''(\omega))^2+\omega^2\tilde\gamma'^2(\omega)}. \label{msv_numerical}
\end{align}
\begin{table}[ht]
\centering
\small
\begin{tabular}{ |>{\centering\arraybackslash}p{3.3cm}|>{\centering\arraybackslash}p{3cm}|>{\centering\arraybackslash}p{2cm}|>{\centering\arraybackslash}p{2cm}|  }
\hline
Relative interaction & Dynamical regime & $ \Delta^2_x(t)$ & $ \Delta^2_v(t)$ \\
\hline
Strong Coupling
 & $t\ll \{\tau_c,\tau,\tau_b\}$  & $ t^2 $ & $ t^2 $  \\
 \textcolor{black}{$\lambda>k$ ($\tau_c<\tau_b$)}  & $\tau_c\ll t\ll\{\tau,\tau_b\}$   & $ t^2 $& $ t^2 $ \\
 & $\tau_c<\tau\ll t\ll \tau_b$   & $ t $& $ \sqrt{t} $ \\
 & $\tau_c<\tau_b<t\lesssim \tau$   & $ t^2 $& $ \sqrt{t} $ \\
 & $\{\tau_c,\tau,\tau_b\}\ll t$ & $ \sqrt{t} $& $ v_s^2 $ \\
 \hline
 Weak Coupling
 & $t\ll \{\tau_b,\tau,\tau_c\}$ & $ t^2 $ & $ t^2 $ \\
$k>\lambda$ ($\tau_c>\tau_b$) & $\tau_b\ll t\ll\{\tau,\tau_c\}$   & $ t^2 $& $ t^2 $ \\
 & $\tau_b<\tau\ll t\ll \tau_c$   & $ \sqrt{t} $& $ \sqrt{t} $ \\
  & $\tau_b,\tau_c<t\lesssim \tau$   & $ t^2 $& $ \sqrt{t} $ \\
 & $\{\tau_b,\tau,\tau_c\}\ll t$ & $ \sqrt{t} $& $ v_s^2 $ \\
\hline
\end{tabular}
 \caption{Comparison of coupling constants, corresponding time scales, and the variation of mean squared displacement and velocity with time.}
    \label{tab:msd_msv}
\end{table}
The above integral is again difficult to solve exactly, and we solve it numerically to understand the behavior in the different temporal regimes.

\subsubsection{Slower bath relaxation \texorpdfstring{$\tau_c<\tau_b$}{}}
\begin{figure}[t]
    \centering
    \includegraphics[width=\linewidth]{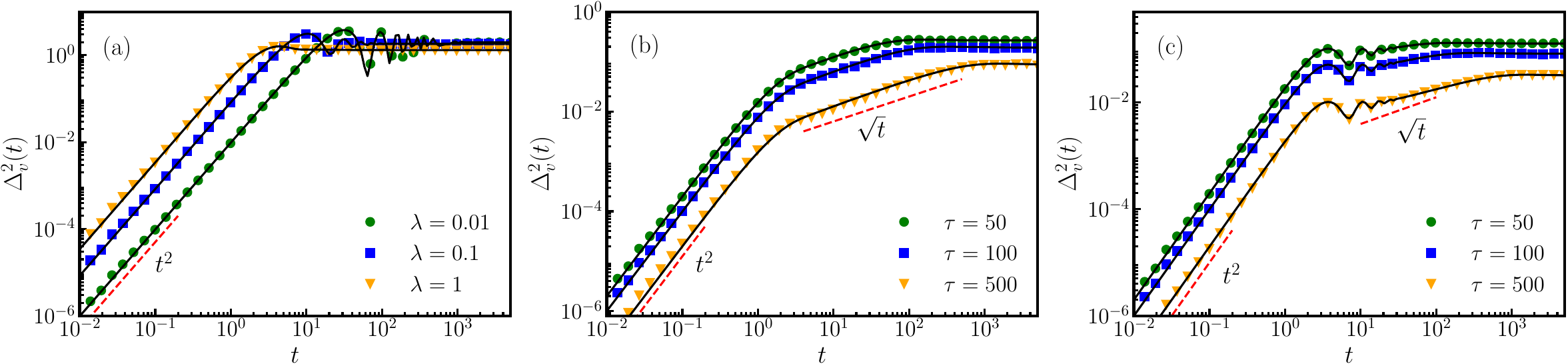}
    \caption{The panels show the MSV of the probe particle attached to an \textcolor{black}{ARB} as a function of time for (a) Small activity: for different values of $\lambda$ and $\nu=1$, $k=0.1$ , $\tau=1$; (b) Strong coupling: MSV for different values of $\tau$ and $\nu=1$, $k=1$, $\lambda=10$. (c) Weak coupling: for different values of $\tau$ and $\nu=1$, $k=10$, $\lambda=1$. The black solid lines correspond to the theoretical prediction Eq.~\eqref{msv_numerical}.\textcolor{black}{Numerical simulations are averaged over $10^4$ trajectories with $\Delta t=5\times 10^{-3}$.}}
    \label{fig:msv}
\end{figure}

If the bath time-scale is the largest time scale, then the MSV shows an initial $\sim t^2$ behavior till $t\simeq \tau_b$, beyond which it saturates to a constant value denoted by $v_s^2$ [see Eq.~\eqref{msv_numerical}] and this is shown in Fig.~\ref{fig:msv}(a). On the other hand, if the active time-scale $\tau$ is the largest, then the MSV shows the same $\sim t^2$ growth till $t\simeq \tau_b$, followed by an intermediate $\sqrt{t}$ behavior for $\tau_b\ll t\ll\tau$, beyond which it reaches the saturation value $v_s^2$ [see Fig.~\ref{fig:msv}(b)].

\subsubsection{Faster bath relaxation \texorpdfstring{$\tau_b<\tau_c$}{}}
In this scenario, when the active time-scale $\tau$ is intermediate between $\tau_b$ and $\tau_c$, then the MSV exhibits a $\sim t^2$ growth for times smaller than $\tau_c$ thereafter saturating to $v_s^2$. In case, when the active time scale is larger than $\tau_c$, the short time $\sim t^2$ value is followed by an intermediate regime $\tau_c\ll t\ll \tau$ with a $\sim \sqrt{t}$ growth, before saturating to constant value $v_s^2$ and this is shown in Fig.~\ref{fig:msv}(c).

\textcolor{black}{In both the above cases, we find a new intemediate regime $t\in (\max\{\tau_b,\tau_c\},\tau)$ for velcoities which shows a $~\sim t^{1/2}$ growth. In this regime the velocity growth is driven only by the thermal fluctuations, and is thus absent for thermal baths\footnote{\textcolor{black}{For thermal bath~[see \ref{thermal_bath}] the noise driving the particles in ARB is delta-correlated, implying that the time-scale corresponding to $\tau$ tends to zero.}}.}

Similar to the position fluctuations, damped oscillations are observed at the onset of the intermediate and late time regimes of the MSV. The oscillations are more pronounced in the weak coupling limit $\lambda<k$, and decreases with an increase in ratio $\lambda/k$ (see Fig.~\ref{fig:msv} (b) and (c)). The behavior of the MSV in the different dynamical regime is summarized in Table~\ref{tab:msd_msv}.

\subsection{Stationary MSV}
\begin{figure}[t]
    \centering
    \includegraphics[width=0.8\linewidth]{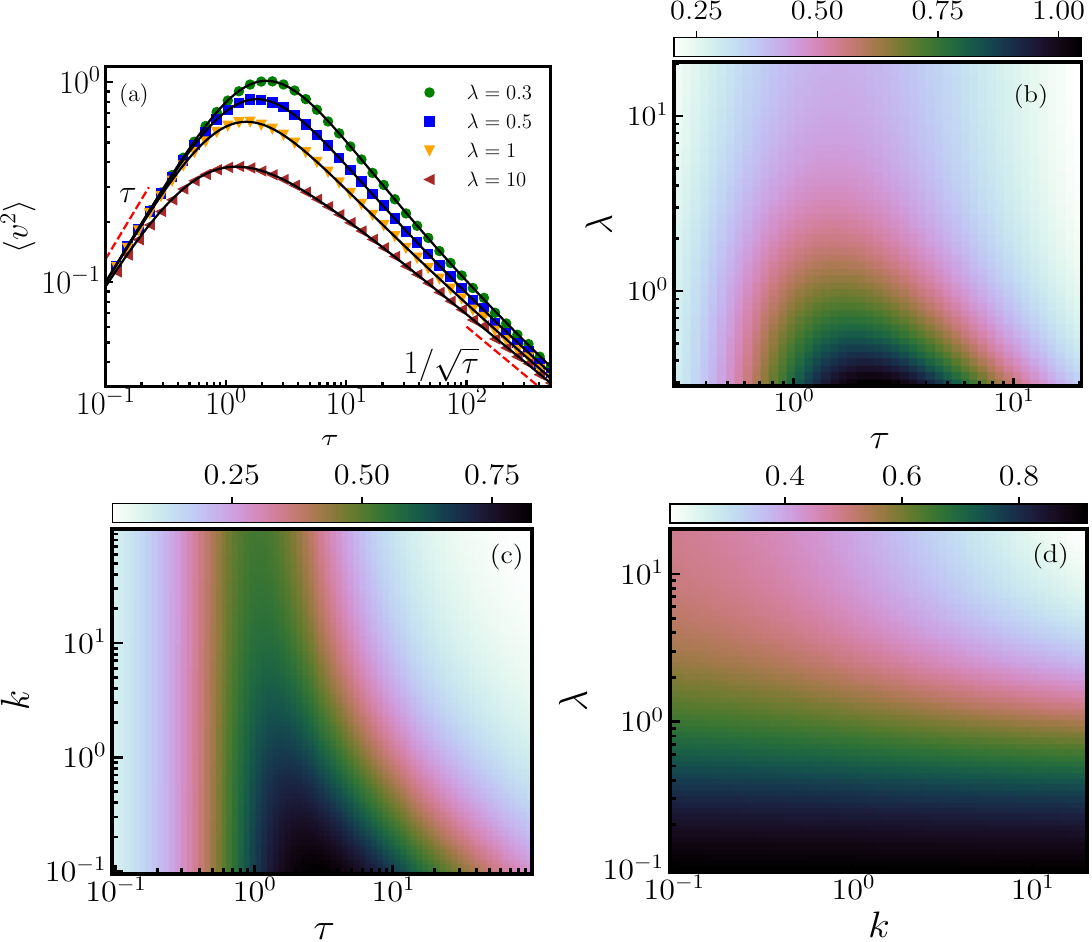}
    \caption{(a) \textcolor{black}{Stationary MSV} $\la v^2 \ra$ as a function of $\tau$ for fixed $k=2$, $m=1$, \textcolor{black}{$\nu=1$} and different set of $\lambda$. The symbols denote the data obtained from numerical simulations and black solid lines corresponds to Eq.~\eqref{eq_ke}. Average kinetic energy of the probe: Color maps of average of $v^2$ in the stationary state with fixed \textcolor{black}{$\nu$=1}, illustrating its variation with different $k$, $\lambda$ and $\tau$. (b)  $\la v^2 \ra$ as a function of $\lambda$ and $\tau$ for $k=1$. (c)  $\la v^2 \ra$ as a function of $k$ and $\tau$ for $\lambda=1$. (d)  $\la v^2 \ra$ as a function of $\lambda$ and $k$ for $\tau=1$. \textcolor{black}{Numerical simulations are averaged in the stationary state 
    with $\Delta t=5\times 10^{-3}$.} }
    \label{fg:vsqrrd}
\end{figure}
At large-times the MSV saturates to a constant value $v_s^2$, which is given by,
\begin{align}
v_s^2=\lim_{t\to\infty}\Delta^2_v(t)=\frac{2}{\nu \pi}\int_0^\infty d\omega \frac{\omega^2\,\tilde h(\omega,\tau)\tilde \gamma'(\omega)}{(-m \omega^2+\omega \tilde\gamma''(\omega))^2+\omega^2\tilde\gamma'^2(\omega)}.\label{eq_ke}
\end{align}
A closed form expression for the above integral is hard to obtain, but we can solve it numerically and understand how the saturation value depends on the time-scales of the problem. It turns out that for a fixed activity, $v_s$ decreases with increase in $k$ (and with $\lambda$) when  $\lambda$ (and respectively $k$) is kept fixed. This is an interesting feature of the active reservoir, which is in sharp contrast to equilibrium case (see~\ref{thermal_bath}), where stationary velocity fluctuations are governed solely by the temperature of the medium $k_BT/(2m)$ irrespective of the microscopic structure of the medium or the tracer medium interaction. This is shown in Fig.~\ref{fg:vsqrrd}(c). 
The dependence of $v_s^2$ on the activity $\tau$ shows more interesting features: $v_s^2$ has a non-monotonic behavior with $\tau$ for fixed $k$ and $\lambda$. The origin of this non-monotonicity lies purely in the active nature of the bath particles. It can be understood by looking at the statistical properties of the bath noise $f_M(t)$ [see Eq.~\eqref{fcorr}]: At very small $\tau$ $f_M(t)$ changes between $\pm v_0$ very fast, and the bath particles essentially behave like thermal particles at an effective temperature $\propto v_0^2\tau$. In fact, in this limit, Eq.~\eqref{eq_ke} reduces to,
\begin{align}
    v_s^2\approx\frac{2\tau}{\nu \pi}\int_0^\infty d\omega \frac{\omega^2\,\tilde \gamma'(\omega)}{(-m \omega^2+\omega \tilde\gamma''(\omega))^2+\omega^2\tilde\gamma'^2(\omega)}.
    \label{smalltau}
\end{align}

On the other hand, in the limit of infinite persistence $\tau\to\infty$, the bath oscillators are very persistent, and the velocities remain fixed at either of $\pm v_0$, causing the velocity fluctuations to be very small--- this accounts for the long $\tau$ decay. Since the integrand in Eq.~\eqref{eq_ke} is dominated by contribution from small $\omega$, the large $\tau$ limit cannot be immediately taken outside the integrand, unlike Eq.~\eqref{smalltau}. However, a numerical evaluation of the integral reveals a $\tau^{-1/2}$ decay in this regime. This is shown in Fig.~\ref{fg:vsqrrd}(a). \textcolor{black}{ For a fixed finite value of $\tau$, $v_s^2$ does not show any non-monotonicity in the $\lambda-k$ plane, and decreases monotonically with both $\lambda$ and $k$, as shown in Fig.~\ref{fg:vsqrrd} (d). }

\section{Autocorrelation and response}\label{autocorrelation}
In equilibrium, the response of an observable due to an external perturbation is proportional to the two-point correlations in the absence of it. In the following we first compute and characterize the stationary velocity correlations $C(t)=\la v(t)v(0)\ra$, and thereafter discuss the linear response of the probe to an external perturbation.

\begin{figure}[t]
    \centering
    \includegraphics[width=0.8\linewidth]{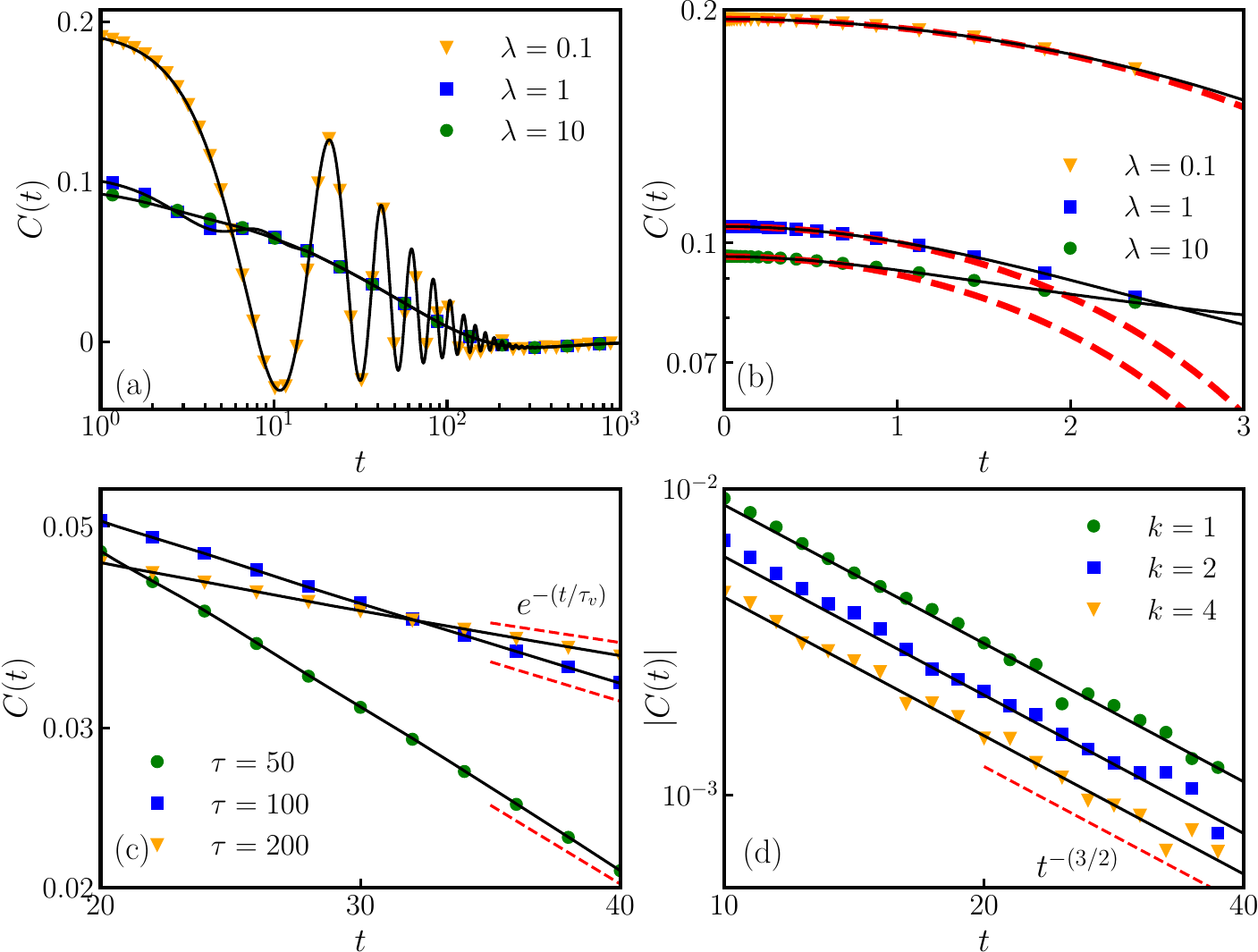}
    \caption{(a) Plot of stationary velocity autocorrelation as a function of time at all time for different value of $\lambda$. (b) Plot of stationary velocity autocorrelation as a function of time in short-time for different value of $\lambda$. The red-dashed lines correspond to the equation $C(t)=C(0)-\alpha t^2/2$. Symbols corresponds (a) and (b) to the result from numerical simulation with the parameters $k=1$, $\nu=1$ and $\tau=100$. (b) Plot of normalized stationary velocity autocorrelation as a function of time in moderate-time regime for different value of $\tau$. Symbols corresponds to the result from numerical simulation with the parameters $k=1$, $\nu=1$ and $\lambda=1$. Red dashed lines correspond exponentially decaying function with decay rate $\tau/2$. (d) Plot of stationary velocity autocorrelation as a function of time for large time for different value of $k$. Parameters used for the numerical simulations are $\lambda=10$, $\nu=1$ and $\tau=1$. Black solid lines in panels (a)-(c) correspond to Eq.~\eqref{time_corr} and the for panel (d), the black solid lines correspond to Eq.~\eqref{time_corr_large}. \textcolor{black}{Numerical simulations are averaged over  $10^5$ [panel (a)] and $10^7$ [panels (b)-(d)]trajectories with $\Delta t=5\times 10^{-4}$.}}
    \label{fig:time_correlation}
\end{figure}

The velocity autocorrelation can be easily computed from Eq.~\eqref{msv_numerical}, and is given as,
\begin{align}
    C(t)=\frac{1}{\nu \pi}\int_0^\infty d\omega \frac{\omega^2\,\tilde h(\omega,\tau)\tilde \gamma'(\omega)\cos{\omega t}}{(-m \omega^2+\omega \tilde\gamma''(\omega))^2+\omega^2\tilde\gamma'^2(\omega)}.\label{time_corr}
\end{align}
The above integral is difficult to obtain exactly, however, we can evaluate it numerically (see Fig.~\ref{fig:time_correlation}(a)) and investigate the different dynamical regimes by extracting the leading order asymptotic limits. For very short-times, the integrand can be expanded as a series in $t$ leading to,
\begin{align}
    C(t)=C(0)-\frac{\alpha}{2} t^2, \quad \text{where~~} \alpha=\frac{1}{\nu \pi}\int_0^\infty d\omega \frac{\omega^4\,\tilde h(\omega,\tau)\tilde \gamma'(\omega)}{(-m \omega^2+\omega \tilde\gamma''(\omega))^2+\omega^2\tilde\gamma'^2(\omega)},
\end{align}
is a decreasing function of $\tau$ for large bath activities. This short-time behavior of the velocity autocorrelation is compared to numerical simulations in Fig.~\ref{fig:time_correlation}(b).

\textcolor{black}{The integrand in \textcolor{black}{Eq.}~\eqref{time_corr} is hard to do it analytically, however, evaluating the integral numerically reveals that, at late-times, the velocity correlations shows an exponential decay $\exp(-t/\tau_v)$ with 
 $\tau_v/\tau=0.5$ [see Fig.~\ref{fig:time_correlation} (c)]. Note that this intermediate exponentially decaying regime is absent for thermal baths which corresponds to $\tau\to0$.}

For very large times, $C(t)$ shows a power-law decay: this can be seen by taking $\omega=z/t$ in Eq.~\eqref{time_corr} and thereafter using the long $t$ asymptotes of $\tilde\gamma(z/t)$,
\begin{align}
    C(t)\approx \frac{\tau }{2 \sqrt{\pi } \sqrt{k \nu ^3}}t^{-3/2}.\label{time_corr_large}
\end{align}
The analytical prediction of velocity autocorrelation $C(t)$ at large time is compared to numerical simulation in Fig~\ref{fig:time_correlation}(d) for different values of $k$ and show excellent agreement.

Let us now discuss the response of the system to an external perturbation $V_{\text{ext}}(x,t)$ which couples linearly to the probe position $V_{\text{ext}}(x,t)=-xf_{\text{ext}}(t)$,
\begin{align}
    m\ddot{x}(t)&=-\int_{-\infty}^t ds \, \Gamma(t-s)\, \dot{x}(s)+\zeta(t)+f_{\text{ext}}(t)
\end{align}
The response of the probe to the external force results in a non-zero mean velocity of the tracer $\la v(t)\ra=\int_0^t dt' R(t-t')f_{\text{ext}}(t')$, where the response function $R(t)$ is expressed conveniently in frequency space as 
\begin{align}
    \tilde R(\omega)=[-im\omega +\tilde\gamma(\omega)]^{-1}
\end{align}
Note that, the response function itself is independent of the activity of the system. This can be attributed to the quadratic nature of tracer-bath, and inter-bath particle couplings. \textcolor{black}{Interestingly, activity independent response functions for tracer in bacterial baths have also been observed in experiments~\cite{turlier2016equilibrium,maggi2017memory}.}

The velocity autocorrelation Eq.~\eqref{time_corr} in the frequency domain can be written in terms of the response function as,
\begin{align}
 \la v(\omega) v(\omega')\ra =\frac{\la \tilde\zeta(\omega)\tilde\zeta(\omega')\ra}{\tilde\gamma'(\omega)^2+(-m\omega+\tilde\gamma''(\omega))^2}=\frac{2\pi}{\nu}\tilde h(\omega,\tau)\tilde R'(\omega)\delta(\omega+\omega').
 \end{align}
\begin{figure}
    \centering
\includegraphics[width=0.85\linewidth]{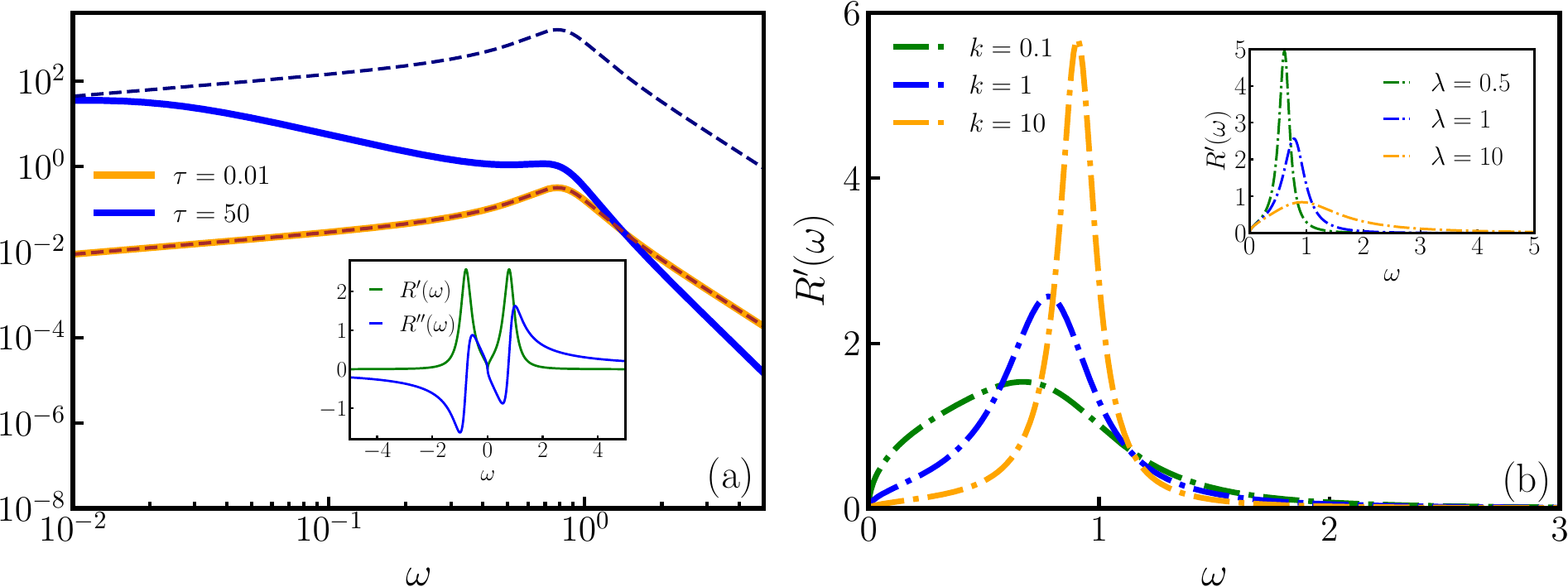}
    \caption{Panel (a) shows the validity and non-validity of the FDT for baths with small and large activities: the solid and dashed lines denote the rhs and lhs of Eq.~\eqref{fdt:1} respectively, for $m=k=\lambda=\nu=1$. The inset shows the real and imaginary parts of $\tilde{R}(\omega)$. Panel (b) shows the real part of the response function for different interaction strengths of the bath oscillators. }
    \label{fig:vel_response}
\end{figure}
The above equation denotes the modified Fluctuation response relation (also called FDT of the first kind~\textcolor{black}{\cite{kubo2012statistical,zwanzigbook,balakrishnan1979fluctuation}}) where, compared to the  well-known equilibrium form, the temperature of the bath is replaced by  the frequency dependent factor $\tilde h(\omega,\tau)$. For small activities $\tau\to 0 $, $\tilde h(\omega,\tau)\approx 2v_0^2\tau$, and the equilibrium form of FDT is restored with the effective temperature $k_BT_{\text{eff}}=v_0^2\tau/\nu$,
\begin{align}
     \la v(\omega) v(\omega')\ra =4\pi k_B T_{\text{eff}}\tilde R'(\omega)\delta(\omega+\omega').\label{fdt:1}
\end{align}
This is illustrated in Fig.~\ref{fig:vel_response}(a) for different values of activity $\tau$.

The real and imaginary parts of the response functions gives us some important information about the system. We discuss them for an oscillatory drive $ f_{\text{ext}}(t)=f_0\cos(\Omega t)
$. Using this drive in the response relation, the mean tracer velocity in the stationary state is given by,
\begin{align}
    \la v(t)\ra=f_0\left(\tilde R'(\Omega)\cos(\Omega t)+\tilde R''(\Omega)\sin(\Omega t)\right)
\end{align}

The rate of work done on the tracer due to the external drive is given by,
\begin{align}
    \frac{dW(t)}{dt}=f_{\text{ext}}(t)\la v(t)\ra.
\end{align}
For the considered oscillatory drive the average rate of work done over one period $T=2\pi/\Omega$ is,
\begin{align}
    \frac{d\bar W}{dt}=\frac{1}{T}\int_0^T \frac{dW(t')}{dt'} dt' =\frac{f_0^2}{2} \tilde R'(\Omega).
\end{align}
The real part of the response function $\tilde R'(\omega)$ is peaked at $|\omega|=\omega^*$. Both $\omega^*$ and the maximum value $\tilde R'(\omega^*)$ increases with an increase in $k$ [as shown in Fig.~\fref{fig:vel_response}(b)], indicating that stronger bath interactions enhance the amount of work that can be extracted from tracers. On the other hand, stronger tracer-bath coupling increases the dissipation of the tracer and hence expectedly the extractable work decreases with $\lambda$, as shown in the inset of Fig.~\ref{fig:vel_response}(b).

\section{Harmonic confinement: Energetics and equipartition theorem}\label{energetics}
In this section, we confine the tracer particle in a harmonic trap, i.e., $U(x)=\mu x^2/2$ in Eq.~\eqref{eq:x}. This typical setup has been used in experiments to investigate equipartition of energy. At long-times, both the velocity and position reach a stationary state and the average potential $\mathcal{P}$ and kinetic $\mathcal{K}$ energies are given by,
\begin{align}
\mathcal{K}=\frac{1}{2}m\la v^2\ra,\quad \mathcal{P}=\frac{1}{2}\mu \la x^2\ra
\end{align}
For a tracer particle coupled to equilibrium baths, equipartition theorem holds and one gets $\mathcal{K}=\mathcal{P}=k_BT/2$. In this section we investigate if such a relation holds for the case the active Rubin bath.

To this end, we note that, the tracer degree of freedom in frequency space is,
\begin{align}
    \tilde x(\omega)=\frac{\tilde\zeta(\omega)}{-m\omega^2+\mu-i\omega\tilde\gamma(\omega)}, \quad \tilde{v} (\omega)=-i\omega \tilde x(\omega).
\end{align}
Using the above solution, the potential energy and the kinetic energy of the tracer can be obtained as,
\begin{align}
    \mathcal{P}&=\frac{\mu \la x^2\ra}{2}=\frac{\mu}{\nu}\int_0^{\infty}\frac{d\omega}{2\pi}\frac{\tilde \gamma'(\omega)\tilde h(\omega,\tau)}{[-m \omega^2+\mu+\omega \tilde\gamma''(\omega)]^2+\omega^2\tilde\gamma'^2(\omega)},\label{pe_active}\\
    \mathcal{K}&=\frac{m \la v^2\ra}{2}=\frac{m}{\nu}\int_0^{\infty}\frac{d\omega}{2\pi}\frac{\omega^2\tilde \gamma'(\omega)\tilde h(\omega,\tau)}{[-m \omega^2+\mu+\omega \tilde\gamma''(\omega)]^2+\omega^2\tilde\gamma'^2(\omega)},\label{ke_active}
\end{align}
where we have used the effective noise correlations Eq.~\eqref{fcorr}. 
An exact closed form expression for the above integrals is difficult to obtain, however, it can be evaluated numerically. 
\begin{figure}[t]
    \centering
    \includegraphics[width=0.9\linewidth]{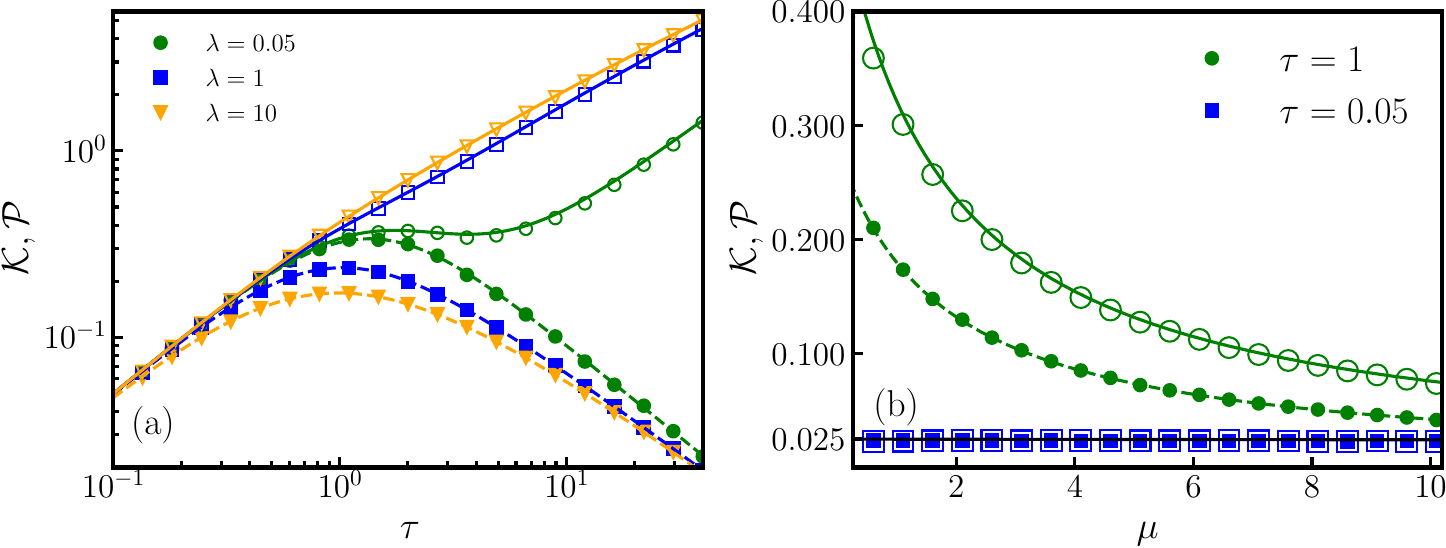}
    \caption{Average kinetic and potential energy of the probe particle in a trap of trap strength $\mu$: (a) Average kinetic and potential energy is plotted as functions of activity of the Rubin bath $\tau$ for fixed $k=1$, $\nu=1$, $\mu=0.5$ and different set of $\lambda$. The filled symbol (open symbols) corresponds to average kinetic energy (potential energy) evaluated using numerical simulation.  (b) Average kinetic and potential energy of the trapped probe is plotted as a function of the trap strength for fixed $k=10$, $\nu=1$, $\lambda=1$ and different set of $\tau$. The solid and dashed lines in both (a) and (b) correspond to the analytic value of average kinetic and potential energy evaluated by Eqs.~\eqref{pe_active} and \eqref{ke_active}. \textcolor{black}{Numerical simulations are averaged in the stationary state 
    with $\Delta t=5\times 10^{-3}$.}}
    \label{fig:energetics}
\end{figure}

\begin{figure}[t]
    \centering
    \includegraphics[width=0.9\linewidth]{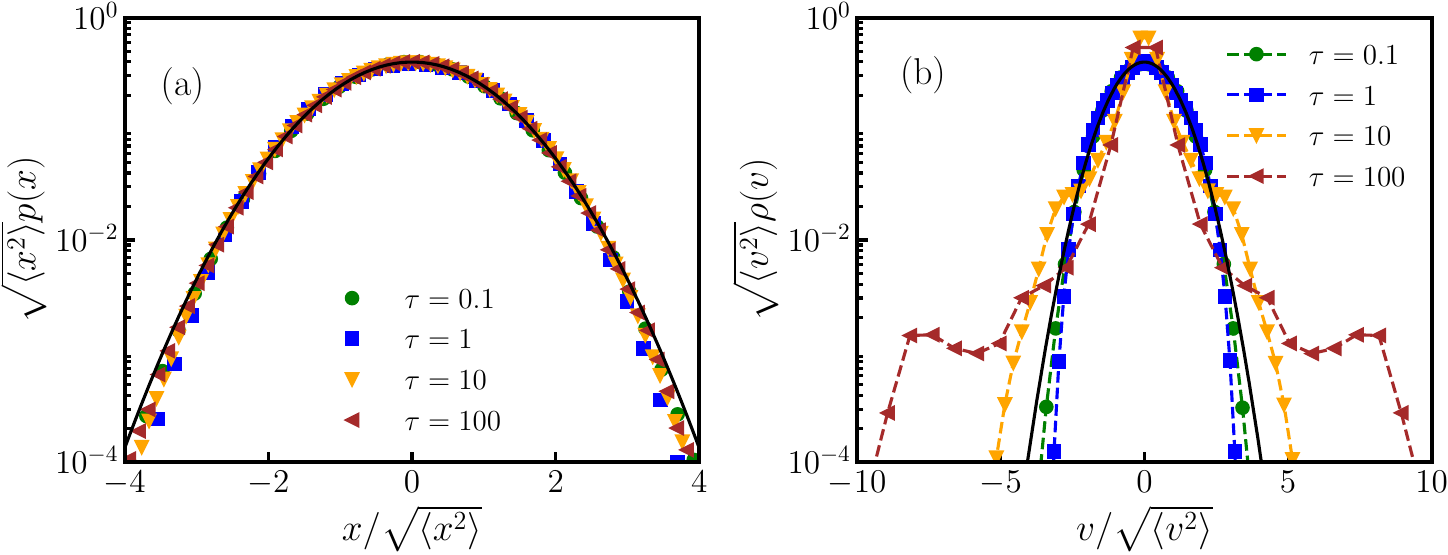}
    \caption{Plot of (a) scaled position distribution and (b) scaled velocity distribution for different activity $\tau$. The symbols correspond to the result we get from numerical simulation with the parameters $k=1$, $\lambda=10$, $\nu=1$ and $\mu=0.5$. The black solid lines correspond to standard Gaussian distribution with zero mean and unit variance. \textcolor{black}{Numerical simulations are averaged in the stationary state 
    with $\Delta t=5\times 10^{-3}$.}}
    \label{fig:dist_compare}
\end{figure}

Let us analyze the integrals in the small activity limit. The integrands in $\mathcal{P}$ and $\mathcal{K}$ decrease as $\sim\omega^{-6}$ and $\sim\omega^{-4}$ for large values of $\omega$; thus the main contribution to the integral comes from small but finite $\omega$ of the integrand. Due to this, for very small activity, i.e., $\tau\to 0$, the dependence on the activity can be extracted simply by putting 
\begin{align}
\tilde h(\omega,\tau)=\frac{2v_0^2\tau}{1+\omega^2\tau^2}\approx 2v_0^2\tau,
\end{align}
thus  yielding a linear
dependence on $\tau$,
\begin{align}
    \mathcal{P}&=2\mu k_B T_\mathrm{eff}\int_0^{\infty}\frac{d\omega}{2\pi}\frac{\tilde \gamma'(\omega)}{[-m \omega^2+\mu+\omega \tilde\gamma''(\omega)]^2+\omega^2\tilde\gamma'^2(\omega)},\label{poten:smalltau}\\
    \mathcal{K}&=2m k_B T_\mathrm{eff}\int_0^{\infty}\frac{d\omega}{2\pi}\frac{\omega^2\tilde \gamma'(\omega)}{[-m \omega^2+\mu+\omega \tilde\gamma''(\omega)]^2+\omega^2\tilde\gamma'^2(\omega)},\label{kinen:smalltau}
\end{align}
where $k_BT_{\text{eff}}=v_0^2\tau/\nu $. Therefore, in the small-activity limit, the passive probe experiences an effective temperature $T_{\text{eff}}$ [see \ref{thermal_bath}] and in this limit, we have an effective equipartition theorem
 \begin{align}
    \lim_{\tau\to 0} \mathcal{K}=\lim_{\tau\to 0}\mathcal{P}=\frac{k_BT_{\text{eff}}}{2}.\label{eff_temp}
 \end{align}
The effective thermal picture in this small activity regime is also consistent with the stationary position $p(x)$ and velocity distributions $\rho(v)$ which are of the Boltzmann form,
 \begin{align}
     p(x)=\frac{1}{\sqrt{\pi k_BT_{\text{eff}}}}\exp\left[-\frac{\mu x^2}{k_BT_{\text{eff}}}\right],\quad \rho(v)=\frac{1}{\sqrt{\pi k_BT_{\text{eff}}}}\exp\left[-\frac{m v^2}{k_BT_{\text{eff}}}\right].
 \end{align}
The stationary position and velocity distributions obtained from numerical simulations are shown in Fig.~\ref{fig:dist_compare}. The Boltzmann show a good agreement in the small activity limit.

For larger values of $\tau$, the integrals of the average kinetic and potential energies no longer simplify to yield an effective temperature structure as in Eq.~\eqref{eff_temp}, and must be evaluated numerically to understand their dependence on activity. As $\tau$ increases, the kinetic and potential energies deviate from each other, leading to a violation of energy equipartition: the potential energy increases monotonically with \(\tau\), while the kinetic energy decreases. This violation is illustrated for different values of the coupling strength $\lambda$ in Fig.~\ref{fig:energetics}(a).  

Since the kinetic energy is essentially the velocity fluctuations, we look at the stationary velocity distribution of the tracer. We find that, the velocity distribution of the tracer undergoes a shape transition—from a single-peaked Gaussian to a multi-peaked structure with an increase in the bath activity. Intuitively, as $\tau \to \infty$, a given force configuration $\{f_1, f_2, \dots, f_M\}$ (see Eq.~\eqref{eom:bath1}) of the active bath persists for a long time, effectively leading to a deterministic equation of motion for the tracer (Eq.~\eqref{eom_tracer}). This results in additional peaks in the velocity distribution and a narrowing of the central peak at $ v = 0 $, consistent with the decreasing kinetic energy at high activities. Although the velocity distribution resembles that of an inertial run-and-tumble particle~\cite{debraj_trapped_inertial_rtp}, the position distribution remains Gaussian for all values of activity, with its width increasing with $\tau$. This can be attributed to the thermodynamically large bath, which allows for a broad range of active force configurations.  

Interestingly, in highly active baths, the average kinetic and potential energies become dependent on the strength of the confining potential $\mu$, as shown in Fig.~\fref{fig:energetics}(b). This behavior contrasts sharply with equilibrium baths, where $\mathcal{K} = \mathcal{P} = k_B T/2$, independent of $\mu$. However, when the bath activity is low, this independence is restored, making the effective temperature description in Eq.~\eqref{eff_temp} consistent.

\textcolor{black}{Experimental evidence for a generalized energy equipartition has been reported for Silica beads immersed in an E. coli suspension~\cite{maggi_activebath}. 
Although our model is not intended to quantitatively replicate the experimental setup of Ref.~\cite{maggi_activebath}, the experimentally accessible regime, where the bath correlation times were much shorter than the other relevant timescales, allowed for an effective temperature description, consistent with our theoretical predictions. Our findings suggest that for longer bath correlation times—achievable in systems like self-propelled colloids or Janus particles—the effective temperature along with energy equipartition may break down. Our results open avenues for experimental validation in regimes not yet explored.}

\section{Conclusion}
In this paper, we study an exact analytically treatable model of a massive tracer coupled to an interacting active bath, modeled by a chain of active oscillators. We find the properties of the force on the tracer due to coupling with the active bath. These can be divided into a non-local dissipation, characterized by a memory kernel, and a stochastic force--- we compute these exactly in the limit of thermodynamically large baths and discuss the modified FDT that emerges from this model. We find that due to the long-time $t^{-1/2}$ decay of the memory kernel, it is not in general possible to obtain an effective Markov description for this model. However, we discuss that in the strong coupling limit (coupling strength much larger than the inter-bath interactions) one could still make a time-scale separation approximation to get an effective Markov description. Thereafter, we compute the mean squared displacement, velocity of the tracer, and find interesting dynamical regimes. We also find the different dynamical regimes of the velocity autocorrelation of the tracer, and find how the equilibrium form of the  linear response function is modified due to the presence of the activity in the bath. We find the response function and a modified response relation for tracers in active baths. The response function is independent of the activity of the bath, which captures experimental observations in bacterial baths. Finally, we place the tracer in a harmonic confinement, and compute the kinetic and potential energies of the tracer. We find that though an effective equipartition of energy exists for small values of bath activity, it breaks down for with an increase in the bath activity. We also find that, in contrast to equilibrium, the both the kinetic and potential energies depend on the strength of the harmonic confinement.

The analysis presented here is exact, enabled by the bilinear coupling between the tracer and the oscillator chain. The resulting expressions remain valid for microscopic models of well-known active particle dynamics, making them broadly applicable to a range of active matter systems. \textcolor{black}{From a thermodynamic perspective, a widely discussed issue is whether the modified fluctuation–response relations allow a direct assessment of energetic dissipation in active media via the Harada–Sasa equality~\cite{harada2006energy}. While this approach correctly captures the dissipation for certain nonequilibrium dynamics~\cite{santra2025brownian}, there are cases where an extension of the standard relation becomes necessary~\cite{harada_1,harada_2}; it would be interesting to see the consequences of activity and the power-law memory in this context.} The predicted modification of energy partitioning could also be tested experimentally in bacterial baths, as in~\cite{maggi_activebath}. Interestingly, the tracer velocity distribution exhibits a non-Gaussian, multi-peaked structure, reminiscent of that found in active gels~\cite{gov_dist}, where a phenomenological model was used. A detailed study of this distribution—its dependence on interaction types and finite bath sizes—would be worthwhile. Further investigations on tracer fluctuations with introduction of disorder in bath couplings or activities would be interesting, and may reveal whether an effective temperature description can emerge for the probe dynamics.

\section{Acknowledgments}
RS acknowledges CSIR Grant No. 09/0575(11358)/2021-EMR-I.

\appendix
\section{Pure system \texorpdfstring{$k=\lambda$}{}}\label{app_pure}
In Ref.~\cite{prsa_rubin}, the detailed derivation of dissipation kernel and noise correlations for the  \textit{pure system} ($k=\lambda$) are given. However, in this section, we provide a brief description about the the main results.

For pure system, one can write the Green's function matrix $\Lambda(\omega)$ of the pure system using Eq.~\eqref{lambda_1} as,
\begin{align}
    \tilde\Lambda^{0}(\omega)=[-i \omega \mathbb{I}-\Phi^0]^{-1},~~\mathrm{with}~~\Phi_{ij}^0=\frac{\lambda}{\nu}(2 \delta_{ij}-\delta_{i~j-1}-\delta_{i~j+1})
\end{align}
The effective probe noise $\zeta(t)$ and the dissipation kernel $\gamma^0(t)$ for this pure system are defined by the following equations,
\begin{align}
 \zeta^0(t) &=\frac{\lambda}{\nu}\int_{-\infty}^t ds\, \Lambda^0_{Mj}(t-s)f_j(s),\\
   \Lambda^0_{MM}(t) &=-\frac{d}{dt}\Gamma^0(t) ~~\text{with } \Gamma^0(t)=\gamma^0(t)\Theta(t).
\end{align}
The dissipation kernel $\gamma^0(t)$ in the time-domain is,
\begin{align}
    \gamma^0(t)=\lambda e^{-\frac{2 \lambda t}{\nu}}\Bigg[ I_0 \Bigg(\frac{2 \lambda t}{\nu}\Bigg)+I_1\Bigg(\frac{2 \lambda t}{\nu}\Bigg)\Bigg]\Theta(t).
\end{align}
where $\Theta(z)$ is the Heaviside-theta function and $I_n(z)$ denotes the $n$-th order modified Bessel function of the first kind. The real and imaginary part of the dissipation kernel of the \textit{pure system} is given by,
\begin{align}
    \tilde{\gamma}^{0}{'}(\omega)=\frac{\nu}{2}\left( \sqrt{\frac{1}{2}+\sqrt{\frac{1}{4}+\frac{4 k^2}{\nu^2\omega^2}}}-1\right),~~ \text{and}~~
     \tilde{\gamma}^{0}{''}(\omega)=\frac{\sqrt{2 }k}{\omega\left(1+\sqrt{1+\frac{16 k^2}{\nu^2 \omega^2}}\right)^{1/2}}.
\end{align}
Here, $\tilde\gamma^0{''}(\omega)$ and $\tilde\gamma ^0{''}(\omega)$ are even and odd function of $\omega$ respectively. The modified Fluctuation dissipation theorem (FDT) for the probe coupled to the \textit{pure system} is given by,
\begin{align}
\la\tilde \zeta^0(\omega)\tilde\zeta^0(\omega')\ra=\frac{2\pi\,\tilde\gamma^0{'}(\omega)}{\nu}\tilde h(\omega,\tau)\delta(\omega+\omega').
\end{align}
The explicit expression of $\tilde h(\omega,\tau)$ is same as the expression mentioned in Eq.~\eqref{noise_1}
\section{Details of numerical simulation}\label{app_numerucal}
\textcolor{black}{To simulate the dynamics of the tracer attached to an active Rubin bath, we first consider the motion of the probe particle and discretize the Langevein equation of the probe in a harmonic trap of strength $\mu$ [see Eq.~\eqref{eq:x}] in time steps of duration $\Delta t$. Therefore, the position and velocity ($x, v$) of the probe can be updated following the velocity-Verlet algorithm to the second order of $\Delta t$ as,
\begin{align}
    x(t+\Delta t)&=x(t)+\Delta t \, v(t)+\frac{\Delta t^2}{2} g(t),\cr
    v(t+\Delta t)&=v(t)+\frac{\Delta t}{2}\Big[g(t+\Delta t)+g(t)\Big].
\end{align}
Here $g(t)$ is the force acting on the probe particle given by,
\begin{align}
    g(t)=-\lambda\Big[x(t)-y_M(t)\Big]-\mu \, x(t).
\end{align}
Similarly, the equation of motion of the $M$ overdamped active oscillators [given in Eqs.~\eqref{eom:bathl} and \eqref{eom:bath1}] can be discretized in time steps of duration $\Delta t$ to first order. The position of the $l$-th oscillator and the $M$-th oscillator are updated as,
\begin{align}
    y_l(t+\Delta t)&=y_l(t)+\frac{k\,\Delta t}{\nu}\Big[y_{l-1}(t)+y_{l+1}(t)-2 y_l(t)\Big]+\frac{ \Delta t}{\nu} \, f_l(t)\quad\forall \{l:1,M-1\},\cr
    y_M(t+\Delta t)&=y_M(t)+\frac{k\,\Delta t}{\nu}\Big[y_{M-1}(t)-2 y_M(t)\Big]+\frac{\lambda \Delta t}{\nu}x(t)-\frac{(\lambda-k)\Delta t}{\nu}y_M(t)+\frac{ \Delta t}{\nu} \, f_M(t).
\end{align}
Finally, the active force acting on the $l$-th oscillator flips its sign from $v_0$ to $-v_0$ (or vice-versa) with rate $(2\tau)^{-1}$.} \textcolor{black}{For all the data presented in this paper, we have fixed the number of the oscillator in the ARB to be $M=512$.}

\section{Thermal bath}\label{thermal_bath}
It can be easily shown that how the observables like kinetic energy or potential energy of the probe particle change when the constituent active particles of the active bath get replace by passive particles. The passive particles are acted upon by thermal noises $\eta_l(t)$ of strength $T$, so the Eq.~\eqref{fcorr} becomes,
\begin{align}
    \la \eta_l(t)\eta_{l'}(0)\ra=2 \nu k_B T \delta(t)\delta_{ll'}.
\end{align}
The effective noise acting on the probe particle because of the thermal bath is given by,
\begin{align}
 \zeta^\mathrm{th}(t) &=\frac{\lambda}{\nu}\int_{-\infty}^t ds\, \Lambda_{Mj}(t-s)\eta_j(s).
\end{align}
The noise correlation in the frequency space is,
\begin{align}
\la\tilde \zeta^\mathrm{th}(\omega)\tilde\zeta^\mathrm{th}(\omega')\ra=4\pi k_BT\tilde\gamma'(\omega)\delta(\omega+\omega').\label{fdt_therma}
\end{align}
From the above equation, it is clear that for the thermal bath the effective noise spectrum and the dissipation kernel are related by the FDT.

\begin{figure}[t]
    \centering
    \includegraphics[width=0.9\linewidth]{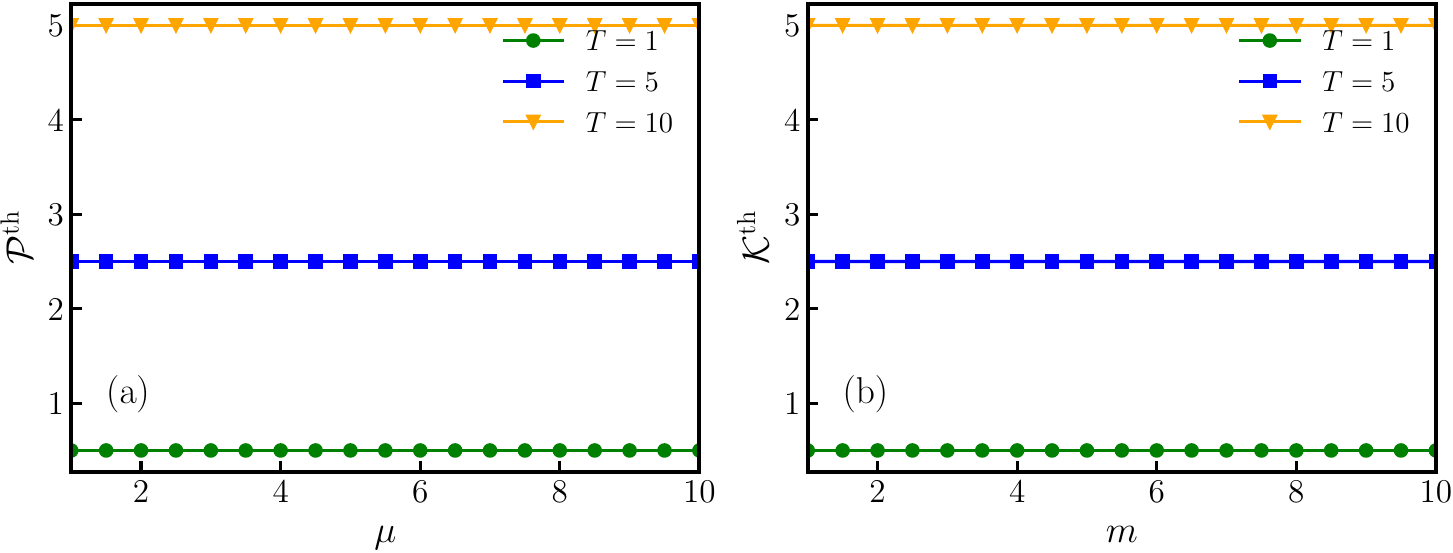}
    \caption{Plot of potential energy (a) and kinetic energy (b) of a tracer particle in a harmonic trap attached to a thermal bath as a function of trap strength $\mu$ and mass of probe $m$. Symbols in correspond to numerical integration of Eqs.~\eqref{thermal_pe} and \eqref{thermal_ke} for fixed $\lambda=10$, $k=3$, $\nu=1$, $m=2$[for(a)], and $\mu=5$[for(b)]. }
    \label{fig:ke_termal}
\end{figure}

Let us write down the potential and the kinetic energy of the tracer attached to the thermal bath can be obtained using Eqs.~\eqref{pe_active}, \eqref{ke_active} and \eqref{fdt_therma} as,
\begin{align}
    \mathcal{P}^\mathrm{th}&=\frac{\mu \la x^2\ra^\mathrm{th}}{2}=2\mu k_BT\int_0^{\infty}\frac{d\omega}{2\pi}\frac{\tilde \gamma'(\omega)}{[-m \omega^2+\mu+\omega \tilde\gamma''(\omega)]^2+\omega^2\tilde\gamma'^2(\omega)}
    ,\label{thermal_pe}\\
    \mathcal{K}^\mathrm{th}&=\frac{m \la v^2\ra^\mathrm{th}}{2}=2mk_B T\int_0^{\infty}\frac{d\omega}{2\pi}\frac{\omega^2\tilde \gamma'(\omega)}{[-m \omega^2+\mu+\omega \tilde\gamma''(\omega)]^2+\omega^2\tilde\gamma'^2(\omega)}
    \label{thermal_ke}.
\end{align}
It is difficult to integrate the above equation analytically. \textcolor{black}{ However, we compute them numerically and find the values to be consistent with the equipartition theorem $\mathcal{P}^\mathrm{th}=\mathcal{K}^\mathrm{th}=k_BT/2$, independent of the system parameters $\lambda,k,\nu, m, \mu$, as expected in equilibrium.}

\bibliographystyle{iopart-num}
\bibliography{ref}
\end{document}